\documentclass[11pt]{article}
\usepackage{graphicx} 

\usepackage{hyperref}

\hypersetup{
    colorlinks=true
}

\usepackage{titlesec}

\titleformat{\section}
  {\normalfont\large\bfseries}{\thesection}{}{}

\titleformat{\subsection}
  {\normalfont\large\bfseries}{\thesubsection}{.1em}{}

\usepackage{multicol}
\usepackage{abstract} 
\renewcommand{\abstractname}{}    

\renewenvironment{abstract}
 {\quotation\small\noindent\rule{\linewidth}{.5pt}\par\smallskip
  {\centering\bfseries\abstractname\par}\medskip}
 {\par\noindent\rule{\linewidth}{.5pt}\endquotation}

\usepackage[a4paper, top=0.55in, left=0.75in, right=0.75in, bottom = .75in]{geometry}
\usepackage{authblk}

\title{Conquering images and the basis of transformative action}

\author[1,2]{Hunter Priniski}

\affil[1]{Department of Psychology, UCLA}
\affil[2]{Slow Technology Laboratory, Designated Technologies, LLC}
\date{}

\begin{document}

\maketitle
\vspace{-44pt}
\begin{abstract}
\noindent Our rapid immersion into online life has made us all ill. Through the generation, personalization, and dissemination of enchanting imagery, artificial technologies commodify the minds and hearts of the masses with nauseating precision and scale. Online networks, artificial intelligence (AI), social media, and digital news feeds fine-tune our beliefs and pursuits by establishing narratives that subdivide and polarize our communities and identities. Meanwhile those commanding these technologies conquer the final frontiers of our interior lives, social relations, earth, and cosmos. In the Attention Economy, \textit{our} agency is restricted and \textit{our} vitality is depleted for \textit{their} narcissistic pursuits and pleasures. Generative AI empowers the forces that homogenize and eradicate life, not through some stupid ``singularity” event, but through devaluing human creativity, labor, and social life. Using a fractured lens, we will examine how narratives and networks influence us on mental, social, and algorithmic levels. We will discuss how atomizing imagery---ideals and pursuits that alienate, rather than invigorate the individual---hijack people's agency to sustain the forces that destroy them. We will discover how empires build digital networks that optimize society and embolden narcissists to enforce social binaries that perpetuate the ceaseless expansion of consumption, exploitation, and hierarchy. Structural hierarchy in the world is reified through hierarchy in our beliefs and thinking. Only by seeing images as images and appreciating the similarity shared by opposing narratives can we facilitate transformative action and break away from the militaristic systems plaguing our lives.
\end{abstract}

\begin{multicols}{2}

\section*{Networks of images of Vitality}

Malevolent networks indoctrinate society with images of Vitality: pictures of self-actualization, success, and interior security; an evolving, evading dream of securing heaven in the earthly world. In reality, Vitality is limitless and imageless. Vitality is Supreme. Vitality exists in the feelings inside your nose, eyes, ears, skin, and muscles. Vitality is shared between \textit{all} bodies and things. When one feels vital, their body and mind is actualized in the here and now. They are in a child-like state of ``flow'' \cite{noauthor_play_1975}, where the self momentarily ceases and creation is inherently transformative. 


Vitality is the source of our attention and agency and exists below our self-image. One's attention is our collective connection, and is ``the elementary human unity... the form of life'' \cite{tiqqun2010introduction}. 

Experiencing attention is effortless but embodying its energy requires constant effort. You can not actualize when focusing on images of actualizing, just as you can't act when you reflect on how to act; and you can't love another when focusing on how they ought to show their love to you. We feel vital when we attend without an objective and act without the purpose of securing life. We love when we do not expect love in return. Love and attention, being Vitality, are beyond the grasp of images and the cycles of desire and expectation.

\textit{A broken definition of images.} Images are our pictures of the world, ourselves, and others. Images capture Vitality by representing it at a given moment of time. An image results from a complex chain of psychological processes
that construct a picture of where attention is directed in the world. For example, when we see an apple on a table, we construct an image of it in our mind. This image is incomplete as we can only see one side of the apple, cannot view its interior space, or comprehend is energy potential, among much else. A multitude of facts exist about the apple that we cannot comprehend in unity, as we can only direct our mind towards a subsection of these facts at a given time. This is one reason why images are incomplete. The other reason images are incomplete is because the world is in constant flux. Change is the one true fact of the exterior world. Images are incomplete because by the very next instant, whatever that image aimed to represent no longer exists as the image encodes it; that thing has become something else entirely. 

Images exist solely in our minds, and are mediated, revised, and disseminated through interlinking hierarchies and networks: emerging within belief systems in individual minds and proliferating through social networks that emerge from groups interacting directly (face-to-face) or through media (face-to-screen). 

Artificial Intelligence, social media, online news media communicate \textit{atomizing images} that convince people their vitality\footnote{This use of vitality is lowercase because it refers to an entity attributable to atomized individuals, and it necessitates a separation or division in Vitality, the indivisible source. vitality is a representation of Vitality, and is thus a mere approximation - an incomplete image of the world.} is limited and depleting, and must be increased by exerting more control over the world. Atomizing imagery establishes narratives that promote individuals either to self-optimize, perform excess labor and material consumption, or desire possessing another's attention and natural resources (e.g., lure a lover with a single-family home and a new Tesla). 

Identifying with images and their pursuit consume us with dissonance as we seek coherence. Images place us in a plane of associations and conditioned expression: the field of memory, categories, and cause-effect relationships. There is no radical action or personal transformation in this lower plane, because actions within it are conditioned on images which construe agency as something that is of the past and dead \cite{weil1997gravity}. One's connection to Vitality is their ability to express their agency, one's ability to move through space and time, and have a causal impact on the world, and their unity transcends conditioned expression: our propensity for agency/creativity exists prior to any form of conditioned learning. Agency is our life-force in the here and now, and images position us a constant step behind our life-force because images are reflections of the past or goals for the future rather than aligning one's body with the present. Being ``out of alignment'' leaves us feeling perpetually incomplete and a victim of our circumstance; seeking alignment inspires many to act with confusion or malicious intent. This is why vicious opportunists succeed in optimizing systems. It is exclusively through the plane of images that empires interface with life by conditioning collective action on atomizing imagery. What is not an image is the truth and is beyond language, but most importantly is the sole force that can lift us from the plane of images and the deathgrip of empire and conquest.\footnote{Any logical proposition, causal explanation, metaphor, or narrative is incomplete as it can only represent a subsection of the world at a given time. While these structures are useful in that they enable practical manipulation of the world, this doesn't imply that they are true, or real, ontologically.}

Images discretize our expressions of love, attention, and agency by quantifying them and setting a goal for the self to obtain more in the future. Images convince people that they can become actualized if they can effectively secure their own connection to the source.\footnote{Images mislead and this is most commonly experienced in our consumption of modern advertising. When someone sees a clothing advertisement featuring happy, good-looking (oftentimes, white) people representing some brand, the implication is that buying clothes by that brand will make the viewer more vital by wearing that brand (i.e., embodying the image defined by the advertisement). This image is misleading because it promises the viewer that they will embody the image if they purchased the clothes being sold. Cognitive psychologists understand this process as a misleading causal attribution: The cause of happiness and being good-looking is sold as the clothes, but the real cause for happiness and being good-looking is elsewhere.} Images position one's own actualization into the past or present, and oftentimes place one's own actualization in competition with the actualization of others. One's success is another's loss. Securing life is a zero sum game. 

Any image is a particular representation of Vitality in the mind, which is cast from a particular viewpoint that ultimately reflects one's beliefs about what Vitality is, how it operates, and how it can be secured. Vitality filtered through our memory is second-order, because it becomes a dying image of itself.  Love dies the moment it is measured. 

Any image of Vitality is second-order because it is an incomplete embodiment of it. You can not measure the immeasurable; just as you cannot  conquer another's attention, or secure life and absolve yourself of death. People can maximize vitality (i.e., second-order Vitality) because they can maximize images which represent discrete quantities. But a life spent refining one's images to maximize vitality cannot replace a single instance experiencing true Vitality. No form of conquest can replace the power of fully experiencing one's attention; no amount of conditioned learning can substitute ``seeing the light''.

Images are cast from a particular reference point, which orients their shape and structure to maintain one's overall \textit{self}-image. Images are structured to sustain a body's life by facilitating actions that lead to their acquisition. Actualizing the self, itself a representation, is to make oneself feel alive or vital through embodying the image occupying thought. Because of the self-orientation of images, acting to secure an image requires self-directed thinking, which commits the mind to measurement, utility, and time. The mind composes these representations into interlinking categories and beliefs, which are governed by cause-effect relationships inferred about the world. Through efficient use of categories and cause-effect relationships, empires work to optimize minds, groups, and cultures so people uphold it through their endless search to increase their own vitality and better secure their connection to the source of life. 

Because seeking images serves one to avoid death and sustain life, seeking an image is based on fear. Therefore, any effort to secure vitality in the future is based on fear. Fear-based pursuits include redefining one's self-image, maximizing intellect or financial wealth, or justifying the expansion of technical expertise and capitalist solutions to solve social problems. These operations work to maximize one's security in the world, to make believe that one can overcome their fundamental fear of death and live forever. During the Covid-19 pandemic, we subdivided our personal lives to follow the mandates of public health experts \cite{hurley2023conspiracist}. Doing so, we slid into Zoom meetings, remote work, and face-to-face antagonism. We are still fighting our way out of the computer mediated bubble of safety we dove into during the pandemic, and we will not feel vital until we free ourselves of this avalanche of online regulation.

When one fears that what they have right now is not enough, they consume more, work harder, or inject plastic into their face to feel more vital. Of course when we need to eat we act to increase our vitality. But disorienting images push people into a constant cycle of seeking new images to settle the self. This evolving search does not lead to any form of actualization, but instead exhausts the pursuer and depletes them of their vitality. The point of our writing is not to castigate pursuits for self-improvement or the truth (I am engaged in these cycles in my own life) but rather to locate the potential for such pursuits to alienate rather than invigorate the pursuer and society at large. We point to these processes using a fractured lens (not pedantic logical arguments or overly-coherent narratives), as to not recommend any solution or cure. Any verbal solution is based on images and is therefore incomplete. The only path to transformation is to see images as images and to act orthogonally\footnote{Orthogonality is a term borrowed from mathematics, and is a generalization of ``being perpendicular''. Whereas two lines forming a right angle in 2-dimensional space are perpendicular, two planes (or hyperplanes for that matter) in N-dimensional space forming a ‘right angle' are called orthogonal. Being orthogonal is acting outside the fold of one's conditioning, it exposes bodies to new spaces of change.} to the mechanisms optimizing our identities and societies. 

Images that optimize our actions, bodies, and identities serve as propaganda for the capitalist ideology because they are effective at imbalancing people. Atomizing imagery makes people desire a continuous sequence of new images, and thus push them into continual consumption: desires are replaced with new desires; problems replaced with new problems; lovers replaced with new lovers. Each new image is sought on the basis of how well it is expected to satisfy the self and bring it security. To obtain the ever changing sequence of images and solve the ever changing sequence of problems, we must ``trim the fat” and optimize our efforts and self-image. To optimize oneself is to do more in less time because the clock is always ticking towards death.\footnote{Because atomizing imagery turns activities that tend to populate one's leisure time like playing sports, consuming entertainment, or simply drinking coffee into objects of optimization, how one experiences and enjoys the present is recast as a mechanism for a more secure future. We live by asking ourselves: Which workouts optimize my body in the eyes of others?  Do I understand this film's narrative as deeply as I should? Does my Starbucks order embody my aspirational aesthetic?}  The amount of vitality the next image brings must be greater than the previous ones, because we continually have less time to secure life. We optimize our life to convince ourselves that we will escape death through another's memory. 

When we seek new images, we are met with a frustration that motivates us to further optimize ourselves, our relationships, and the world. We optimize to command exterior spaces, sustain the self, and resolve the frustration of a noisy interior. We act to align attention with desire (to force what is true into what we want to be true), but desire is incompatible with attention as the two can never be brought together. Whereas objects of attention are immediate and of the present moment, objects of desire are separated from the present moment by a time-interval. Whether we're nostalgic for the past or in conquest of the future, the time-interval induced by desiring images separates us from ourselves, one another, and the world.\footnote{While desiring something doesn't inherently imply that one is alienated, as desiring food is a necessary component of securing vitality and life, the time-interval induced by desiring future or past world states allows for alienation to take root. Because in the moment of desire our mind exists outside of the here and now, and we are separated from the one true source of our attention and agency; we are out of step with Vitality.}

A time-interval separates attention from the realization of any image, and this separation sustains an imbalance within that requires calculated intervention to secure. Efforts to resolve the imbalance between attention and an image orients people's actions and identities towards upholding the very hegemonic systems that cause this imbalance. People believe the next image generated by the system will save them if they're lucky enough to secure it. However, securing the next image will only create more chaos and confusion because of the immateriality of images. No person, place, or thing can be completely secured.

Desiring to increase one's connection to the source is as incoherent as a body desiring life. By imposing this incoherence, atomizing imagery consumes our attention with thinking, reflections, and time; thereby infecting bodies with a pathological desire for life.\footnote{Development towards life fetishizes life because life cannot be possessed. We can only fetishize life when experiencing it through images, which is heavily exploited by digital media, particularly social media and digital pornography. According to \cite{tiqqun2012toyg}: ``The deception of porn is that it claims to represent the obscene, making visible the point at which all representation evaporates. In reality, any family meal, any managerial meeting is more obscene than a facial ejaculation.” Why else would the front page of PornHub optimize itself to be inundated by incest fantasy, other than because people are in an evolving search of progressing images and mounting desire? } Desiring life orients the self to develop itself and surroundings through self-directed interventions, which optimize one's relation to others and the world, and develop endlessly towards homogeneity and rigid binaries.\footnote{Since the Enlightenment, linear progress has been the schema of historical narratives embraced and promoted by most Westerners. Societies are seen as developing from agrarian (simple) to modern (sophisticated), into postmodern (more sophisticated), into post-postmodern (even more sophisticated!); the individual self is understood as a rise from nothing into something (the silly schema of the American Dream may be the best known of these stories). Growth is the fundamental axiom of Western narratives and \textit{a priori} justifies senseless technological development and capitalist consumption. The next cure is inherently better \textit{because} it's the next thing. But linear progress strips the world of its life and strips us of our life. It increases the efficiency of exploitation of most people (outside the lucky, generally white-cis few) and will continue on doing so until the system collapses in on itself.}  As long as people and societies continue to optimize themselves, we will continue to bake the earth, exploit human labor, and disenfranchise identities on the other side of the binary. All ultimately in conquest of the next image defined for itself. The unbalancing act induced by atomizing imagery can only be resolved with complete attention towards seeing the images we consume as images, and embracing actions that deconstruct, rather than uphold, the binaries they instill.

The over-bloated information economy, militant police forces, sprawling distribution networks, advertising agencies, elite academic institutions, elementary school classrooms, the main-stream media, online social networks, art markets all operate to force submission to their system and convince the masses that capitalism's solutions can solve the mounting technological and social problems it creates. Much of the hysteria at the present moment stems from the fact that artificial technologies have progressed to the point of sophistication, in which they can produce images at an exponentially accelerating scale. Artificial technologies consisting of, but not limited to, social media, digital news, content recommendation algorithms, generative Artificial Intelligence, online shopping, predictive policing software, streaming services, rideshare companies, app technology are incredibly efficient image-generating machines, and are the weapons of choice among those seeking to maintain their wealth, power, and influence over the masses. Large statistical models trained on data tracking our every move are fine-tuned to predict our momentary desires, which ultimately reflect our drive for (biological ao social) survival. These algorithms disseminate (and with the advent of generative AI, now produce) images curated to continually imbalance people, so people strive to situate themselves at the top of their echochamber and step on the heads and hearts of everyone else to get there.

The deluge of images is optimized to pull us apart at our core, and is toxifying nearly all corners of life. Widespread addition to partisan politics and social media has turned American politics into a circus mirroring reality television, all the while feeding the public highly isolating narratives that breed bickering, polarization, and identity politics. Social media produces images of an ideal life that leads people to commodify every corner of their personal lives. Online dating kills romantic intimacy in a sea of swiping through reductionist profiles. Military-grade surveillance systems produce images of interior safety and control, but are used to closely track our most intimate patterns and justify brutally policing minorities on statistical grounds. Generative AI models are paraded for their ability to reduce human error and produce media and content at an increasing scale, but will result in removing the human element from many more of our basic interactions. Artificial technologies have been shoveled onto the public with minimal regard for how they will impact people's lives and societies, and the rest of us are left to suffer from the consequences of their conquest. 

Despite capitalist structures being the very cause of our anguish, the images they produce convince many to structure their lives around the maintenance of these mounting systems; driving people to uphold them in order to secure their image of the optimal, ideal life. Why else would finding a career be synonymous with finding one's self within the sea of postmodern monotony? As long as we pay attention to the atomizing images these malevolent networks produce and define our lives around obtaining them, we will only worsen our disintegration and sustain their conquest for profit and power. Seeking the next image produced by the system will only increase our hysteria and confusion, as new images are cast from previous ones, and are noisier copies of earlier desires and pursuits. Entropy snowballs, and noisier images enmesh us in increasingly noisy computations which take us farther from the truth and our collective vitality, which is unearthed through shared attention. We must let go of the search that has defined our lives to realize the power of shared attention as the mechanism for transformative action.

\section*{The separating force of images and narratives}
Atomizing imagery sustains self- and societal development by exploiting the separation between attention and desire. Images define the boundaries of our identities and social groups, our historical narratives, and shape the physical environment. They correspondingly enable us to variably ward off and embrace beliefs about what we think we are, with whom we belong, what happened in our past, and what is out there to conquer in the future.  

Images reinforce a separation within our interiors and seeking them sustains a loss of equilibrium that necessitates the use of force to resolve. In the pursuit of security, we strive to balance ourselves, groups, and states by warding-off current images only to embrace the next one: a more niche aesthetic, a more radical political ideology, a more powerful weapon (whether it's a word or a gun), a higher paying job, gallery representation, a Nature publication, an influencer girlfriend. Images motivate us to take control of the world and others to possess the stillness of a secure interior. However, each image we desire does not settle us and only perpetuates our separation.

When we identify with an image and want to obtain it as our own (i.e., to recognize ourselves in that image), we are insecure and desire to secure ourselves through embodying the sought after image. Obtaining an image requires calculated effort, and securing it can only happen in the future or past. Because there is a distance between now and the desired future or longed-for past, a time-interval separates oneself from all images. Thought works to collapse the interval between the present and desired image by optimizing actions (life) through efficient use of images and language (communication of images). Images constitute our memory and influence our beliefs as our memory and beliefs influence the images we pursue. Seeking images that optimize our lives are therefore bound to optimize our societies. 
Images carve the world into categories that enter into cause-effect relationships. Categories discretize a continuous world. Saying something belongs to a certain category requires saying other things do not belong to that category (barring practically trivial categories like ‘the set of all things'). Cause-effect relationships represent how categories should behave and what you can expect from their behavior. They also hold between categories, which define their boundaries of application. Cause-effect relationships describe how instances of some category bring about instances of another category by describing how categories change in time. Categories and causal relations are two sides of the same coin; they are the media of our memory.


Grouping things according to what they are and what they are not is a consequence of using categories and cause-effect relationships to carve up the world. After all, a category would be of no use if it failed to make any discrimination between members and non-members. This discriminative capacity of categories and, by extension, the cause-effect relationships they participate in is what gives images their separating force. If we have an image of ourselves in the present moment that diverges from the image we desire, we orient our attention to align our actions to embody the sought after image. Desiring to embody an image is synonymous to seeking category membership, to be one with the category. Categories provide the rigid boundaries that we hope to feel secure within, as they define for us and others who we are and how we should behave.   

The separating force of images propels bodies through space and time, and situates the self over the other. Shared category membership brings things closer together; being on different sides of a boundary makes things move against one another (just think about the dynamics of nationalism and borders; the modular family structure and subdivision of space in the American suburbs). Categories structure our relationships across all dimensions: the interior and exterior world; the self and the other; the family unit; the nation-state; one's social categories (race, gender, political affiliation). Categories also impose morality on its members and non-members: the former constitute the good and the latter the bad. The morality that categories impose functions to justify the use of force to subjugate and control out-group members (at least those caste as out-group to the hegemony). Empires sustain themselves through people's categories by developing social networks and systems, actors, and technologies that produce images that separate people, and motivate the use of force to control those that are outside the boundaries of one's categories. 

Cause-effect relationships and, by extension, the categories that they relate can facilitate capitalist activity and conquest by enabling effective intervention, possession, and control. We build causal models of our environment to command the world and other people to sustain our own vitality. All biological organisms represent the world to some degree to sustain their life, but the categories and causes and effects humans infer are much more flexible than those used by non-human animals. This flexibility makes the images that orient our attention and constitute our desires more open to revision, reinterpretation, and recombination. The human mind has a certain proclivity for image-seeking, which is exploited by atomizing imagery to open our identities, relationships, and groups to the pressures of capitalist production. Crucially, the flexibility of our representations ensures that different people are oriented toward different, often conflicting goals. This is why strongly-held beliefs about conflicting images lead to depersonalization and polarization, and violence is justified by the state as a moral imperative. To escape the dissonance of conflicting images, the mind searches for a new image to act as a cure; often in the form of increased technical expertise, a more optimized police state, or a more radical political ideology. However, it is through the perpetual separation brought on by image seeking that empires and conquest manifests in the world; we will continue to fall prey to isolation as long as we depend on others to provide us  cures. This is a fundamental consequence of existing in the world for the purpose to sustain one's images. 

The separating force of images is most apparent in how we think about and relate to other people. We construct images of other's minds and intentions in order to predict their actions and preferences. We construct images of what they spend their attention on, in order to manipulate them to spend it on us. Images of another's attention are ultimately constrained by our beliefs about how we can better secure their vitality for our own gain. When understanding love on the basis of images, relationships constitute securing another person in our image of ourselves. Sometimes one may briefly break through the possessive frame (generally on first dates, honeymoons, or while ``making love”), and enjoy the pleasure of experiencing another with complete attention and silence: without their personal images of one another obscuring the lens. Those brief moments are sublime and are outside space and time. The infinite distance separating one from another in memory is completely collapsed by the stillness of love, which is experienced only through shared and reflexive attention. Love is ephemeral, as attention is ephemeral, but memory works to construct an image of love to secure the cause of its stillness (i.e., the ``loved-one”) in the future. (Categories serve the same function when discussing and thinking about love, as evident when people talk about their ``type” when dating.) When experiencing love based on our image of another, we're drawn to secure them to sustain our personal security. 
However, images of love makes us believe that security is outside of ourselves, and thus strips us of our own agency. Our life is reoriented toward securing and desiring another, rather than simply experiencing shared attention, and appreciating another's unique expression. Desiring to secure love results in pangs of fear and anxiety because the desired individual can never be secured: what we hold onto is a dying image of the referent which only partially or even erroneously captures the experience we seek. Consequently, mistaking an image for the other degrades the self at the expense of the other. And the only way to radically transform ourselves and our societies, is to embrace the source of our own agency in our daily lives. This is only achievable through not experiencing one's life though the maintenance of their self-image. This only follows from not positioning one love over others, and not placing oneself over others. Atomizing imagery inspires possessive love because we love with the intent to position our self-image over all else, and allow for the commodification of the most personal dimensions of our being. 

Atomizing imagery expresses categories and cause-effect relationships that embolden the self at the expense of the other, because self-oriented thinking is necessary for economic growth and empires to flourish. Senseless economic activity is based on images of oneself securing themselves over others in the future: I am not this thing now, but through effective intervention (hard work, causal manipulation, securing another), I will become some other, more desirable thing. I will obtain a new categorization for myself in my own eyes and in the eyes of others. And my life will tell a story that others will remember when they think about me. This action right here, and those I plan to execute in the future, will help bring about (cause) this dream, and I will endure forever. These self-interested causal thinkers are driven to position themselves over others. When individuals position themselves over others, they position their images and groups over others, and their state over others. Hierarchy in the world constrains and emerges from hierarchy in thought. Hierarchy in thought emerges from the separating force of images and the binary inherited by representing the world with categories and cause-effect relationships. What is and isn't defines what is good (the self - one's interior) and bad (the other - one's exterior).

The full power of images manifests through the use of narratives and language, which structure and relate images within and across bodies. Narratives are higher-order structures emergent in a system of images, which contextualize and enrich individual images by placing them in relation to other images in the system. Narratives are interlinking networks of images, and structure the category and causal knowledge in our belief systems. The narratives we endorse constitute the way we understand the world and are highly sensitive to the narratives endorsed by our peers.\footnote{Members of competing social groups believe competing narratives. This is seen concretely in the recent anti/pro science disagreements surrounding climate change and the Covid-19 pandemic; historically through opposing religious groups and explanatory narratives; disagreements about political action and cures for social ills. We live in a society where disagreements are no longer about issues, but rather identities. Narratives categorize the other to make us believe that our peers are the causes of our problem, not those at the helms of the image generating machines. We must see the narratives we consume as constructions designed to pull us apart. Disowning the narratives fed to us is a crucial step towards transformative change.} They structure people into groups as a function of the content and relationships between their shared images; people are seen as coherent with one another on the basis of their endorsed narratives. 

Language emerges from two or more bodies acknowledging their shared attention and desiring to communicate their personal images and narratives with each other. Language is more general than spoken and written words, language consists in any form of communication. A single gesture can communicate truths words can never capture. Language is a shared convention that bridges and aligns interior spaces. 

Language aligns bodies on the basis of the content and the relations between their images and narratives. Language allows for the progression of knowledge, revising of individual and collective beliefs, societal development, and tool building. The human proclivity for using language as a tool allows for effective collective problem solving skills that rapidly accelerates technological and scientific advancements. The rapid pace of our technological development allows societies to grow, and grow, and grow, and grow… As societies grow, they increase their need for efficiency and to optimize their constituents to maintain homeostasis. Opportunists can leverage the optimizing power of language to cohere minds and groups around centralizing and polarizing narratives, which facilitate collective action and conflict. 

Narratives that spread far and wide in social networks are optimized to do so, because they can efficiently assimilate into the maximum number of people's belief systems. Certain narrative structures allow for homogeneity in belief systems, and are thus effective tools for spreading political and religious propaganda and curating devoted followers. Narratives can make controlling groups easier as their shared beliefs form a predictable monolith that is easy to manipulate and control (i.e., ‘controlling the narrative'). Disseminating disorienting narratives is the goal of propaganda, misinformation campaigns, and identity politics. Online social networks and media make it easier to spread disorienting narratives than ever before.  

Elites weaponize narratives and exploit the separating force of images to divide and conquer people. Conflicting narratives foster competing groups and hijacks the attention of those consumed with the dissonance. Psychotic opportunists disseminate narratives designed to make their supporters believe that their groups must conquer the other side, either sexually, politically, economically, or militaristically, to sustain their own life and those that they love; more precisely, members of one's groups and categories: family-unit, neighborhood, political party, race, religion, nationality. 

Conflict resulting from narrative disagreements is a key source of systemic and generational violence. Religious wars, political conflicts, military conquest, over-policing of minorities, immigrants, and working classes, the dehumanization of queer people, are justified by destroyers through the use of spiritual, scientific, and political narratives. As long as we ``pay attention” to their images, and define ourselves through their narratives, we will continue acting in their system: replacing our idols with new ideals in accelerating succession. Seeking images produced within their system leaves us vulnerable to their conquest as they bake the earth and seas and dislocate, homogenize, and atomize whatever stands in their way. We must awaken from their sinister phantasmagoria if we want self-determination and agency over our lives. Instead of succumbing to the separating force of images, those divided by conflicting narratives must recognize what is shared between any narrative - the humanity of the individual - and ultimately recognize themselves in each other's narratives. While this mutual intelligibility enables narratives to conflict it also enables narratives to cohere; any resulting mutual self-recognition is a first step toward mutual self-determination.

\section*{Commodifying oneself for another's consumption}

Atomizing imagery structures people's interior lives around commodification, accounting, and control. Seeking progressive images requires optimizing one's connection to other people and the world. Optimization and development enables more efficient forms of conquest, and allows people and societies to secure interiors and dominate exteriors with increased precision. 

In conquest for a higher position (self-security at the level of the individual and social-security at the level of society), atomizing imagery drives people to manipulate and define their lives for consumption. Whether it's a rising-star scientist at a top research institution, or the girl from your hometown declaring her unwavering love for her nth boyfriend on Instagram, those who believe in and produce atomizing imagery full-heartedly experience their life through their self-image and therefore support the maintenance of hierarchy, state, and empires. Their world becomes a marketplace where the totality of their relationships with others is reduced to a zero-sum game, and another's attention must be secured to increase one's own. Life becomes a selfish pursuit of a highly curated identity through which people attempt to distinguish themselves against the monotonous cacophony of postmodern life.

Living in relation to images pulls the self out of its sensing body and into states of reflection, optimization, and prediction. Images promise us that we'll become something great if we work a bit harder, a bit longer. But experiencing life in relation to images and material manipulation is a constant mirage over the horizon. You'll \#RiseAndGrind endlessly towards the image you sought for yourself and shared for others to consume, and you'll grow very thirsty forgetting how to breathe. Representing your life for another's consumption makes you move outside the instant and positions you in a space with endless distance spawned by the separating force of categories and cause-effect relationships. Confusing the map for the territory, people lose touch with their own life and are perpetually imbalanced. As confusion grows, the neural pathways of dissociation are strengthened, and the body learns to believe more strongly that it can only become vital through securing another atomizing image or by conquering another's life.

Escaping time, one may realize the image they are currently latched onto dissolves once obtained. Deflated, most re-enter time to inflate themself by moving on to a new image. But the next image, like the previous, only distracts us from the driver of all possessive thinking and action: Fear of death and non-existence, the forever-loss of one's images. Images cannot conquer death because they are of time, and pursuing their parasitic trace separates one's mind from Vitality; thus perusing them sustains our fear of death, never resolves it. Images cannot save us from our memories because they are composed of our memories. 

When we search for life, we act to define ourselves. We hope to encode our image in another's mind and secure their attention to increase our own vitality. Acting to define ourselves for others' consumption is living through images, and leads to isolation. Images only approximate what they represent, and are by no means complete versions of what they represent. When you secure another's attention, you only increase your vitality (i.e., second-order Vitality) because you secure an image that merely approximates another's attention (a representation of their attention and vitality in your mind). A life spent pursuing another's attention is a life spent in vain. Lived experiences mediated by images merely approximates living because one's expression ultimately depends on images of others. The word apple has no flavor, and your images do not embody your lived experience. But why must we live for our images? 
Images can not embody life because images construe life as something either held in the past (nostalgia) or to be obtained in the future (conquest); they render static what is dynamic and crystallize what is fluid. In reality, life is right here and is outside time, images, and the conquest of optimization. We live collectively through our shared attention and images separate us from this simple fact. By separating us from this fact, images imbalance us and lead us to develop ourselves and the world. In a continual conquest of others in search for ourselves. We live for our images when we stop smelling the sounds.
In development, one understands their life through their self-narrative. They act to make a noise inside another, and impose their narrative on others. Atomizing imagery optimizes narratives so those embodying that narrative arrive at prosperity. Each narrative structured to optimize life shares the same fundamental point: the necessity of a secure interior through the use of force. 

Self-narratives intoxicate people with stories of themselves, and lead people to only care about others through the development of their own image and narrative. Consumed with our stories, we succumb to the fantasy that our images are all that exist, and work to invert the causal dependency between image and referent: Whereas images typically depend on the reality they represent, we can instead work to impose our images on others in a way that might alter their behavior and the world they occupy.  This phenomena results in a looping effects of human kinds \cite{hacking1996looping}. To see our will reified in the world outside our heads, our unique causal impact manifesting in the world, historical narrative, minds of others, we intervene (i.e., act) to impose our personal reality. 

Believing one's stories also literally breeds a reflective anxiety characterized by one's inner-monologue ringing continuously inside their head. How can I secure the hottest lover? How can I live the most ecstatic experiences? How can I convince others I've obtained the very best and arrived? Interior calculations create a lot of noise and confusion. They mount as one seeks increasingly complex images for oneself. How can I maximize engagement with my image? How can I let others know mine's a ``life worth living''? How can I become a BOSS INFLUENCER? 

The self wants to be the object of another's desire, because the power of one's self-image depends on maximizing quantities through images. The more images of the self circulating minds, television screens, billboards, book jackets, social media,,, the more attention one's image secures and the more vital one's image becomes. Through the production of imagery, empires make people believe that other people attending to their self-image can sustain them beyond time and help them escape death.  People believe they can exist after death via their ‘legacy' or, more broadly, through memory of their image.\footnote{The body seizing to represent itself is first-order death. Death of itself in its own memory. Second-order death is death of the self in another's memory. There is only one first-order death, but there are potentially infinite second-order deaths. When people live to become famous, they are seeking second-order immortality. When people live to experience the sights and sounds, they have no concept of second-order immortality intruding on their being. After our biological death when we are decomposing in the ground, which life was better ``spent''?} Why have so many men tried to conquer the world and the minds of others? Why is being an influencer the top career choice for Zoomers? Because we all want to be immortal through images and be absolved of death. Being the object of another's attention allows the self to suck the other's energy and life-force, because directing one's attention is one's life-force. But too much time spent seeking the attention of others will make you feel very, very empty and become very, very mid. 

The internet has allowed for the conquest of more attention than ever before. Before online social networks, one's social network consisted of their immediate friends, family, neighborhood, and town or city. Communities were contained by space and face-to-face interaction. Digital communities collapse and thus transcend space, linking anyone to anyone else in the world. On social media the aggrandized self moves onwards and upwards by expanding its reach through collapsing space in online networks.\footnote{These algorithmic pressures and the images they produce are redefining the fabric of our shared spaces and expediting the flow of gentrification in cities spanning the United States \cite{banks2022attention}.} Online networks allow for rapid dissemination of one's mounting image, to conquer the interior lives of others at increasing scale. The spacelessness of digital networks allows us to conquer minds far away, and seek new magnitudes of second-order vitality. 

Isolation is increased by our deep immersion in online spaces, which encourages our societal dependence on artificial expansion and the consumption of images. On social media, measurements of self-worth are heavily dependent on another's evaluation of you, and is operationalized by how others engage with the identity you curate and disseminate. Social media quantifies attention precisely with engagement metrics (shares, likes, comments, follower counts), which serve as a measure of life and social influence (attention control) in the online world. Attention is immeasurable however, and transcends representation. Seeking images of other's attention only misdirects us towards second-order vitality, which increasingly becomes a mockery of Vitality.

Because social media leads to an expansion of self-worth calculations across a broader network of people, and encourages a greater need to internally optimize one's identity for assimilation and domination, social networks can only accelerate the self- and social-deterioration caused by capitalist thinking. This trend is exacerbated by Artificial Intelligence, which predicts user preferences to more effectively recommend content and new images for people to seek. Artificial Intelligence allows for greater fine-tuning of people's self images, and will lead to more effective extraction of attention and more rapid economic development. Far from a utopian future, Artificial Intelligence will only deliver highly personalized routes to impersonalization.
In digital networks, all aspects of connection and community are forms of commodity exchange, because digital relationships are based on images that only approximate human relationships. We have no opportunity for shared attention when we attend to one another exclusively through our images. Such artificial relationships become a perversion of human relationships as they become more deeply  ‘optimized', a process that enables the simultaneous extraction of more and more wealth and the evacuation of more and more second-order vitality. 

An empty self is a powerful commodity, as its imbalance drives consumption. When we seek higher meaning online we continue to re-identify within ever-changing consumer trends, new images re-caste from the old. Re-aestheticizing one's image turns the self into a commodity, as it acts as a force for more consumption: coffee tagged with the Starbucks logo, new clothing styles in an accelerating fast fashion cycle, pictures of bae on his father's boat. Recasting one's online identity through the production of images commodifies one's life as atomizing imagery, as the content generation encourages further consumption. We convince ourselves we're approaching immortality as we receive more likes and engagement, but we're helping keep the beast that kills us alive.

Through redefining and optimizing one's image for consumption, people commodify themselves and drive capitalist behavior. Self-commodification is done in a variety of ways: producing content for social media companies, articulating one's ‘love' over another\footnote{When love is understood as shared attention, then expressing love through language defines one relationship in contrast to the totality of all relationship. Loving one over another (as a word implies) inherits the separating force of categories, as a boundary around the loved is carved out from the world to signal what is loved, and thus what is not loved. Loving through one's categories is a tool of conquest as it can only serve to secure our self-image through another's attention.}, defining one's self-narrative around their career. Social media allows people to optimize their self-image for mass consumption, and as people adapt their personal experiences so that they can be more effectively mined for content generation, work consumes more and more of leisure time. People become more and more alienated from themselves and from one another. As one optimizes their image, they hope to conquer more people's attention and deplete them of their vitality. Identities are optimized to be easily digestible and to thus spread far and wide (this is why Instagram and TikTok breed mid identities and relationships). The farther one's image spreads, the more vital they hope to feel by securing more people's attention. If one's brand succeeds, they may transcend corporeal mortality (at least through representation in the minds of others). People reappraise their self-image both to assess how well they are representing an idealized (optimized) self for others' consumption and to convince them that they are leading a meaningful life and have something valuable to say.\footnote{These motivations transcend social media as well and have existed well before the internet. How often has an artist or poet pushed themselves through anguish to construct a captivating self-narrative to embolden their work with more meaning and artificial context? How often does an academic clutter their writing with esoteric jargon and overly-technical concepts to make it appear as though they've uncovered some deep truth?}

When optimizing their image to compel others' consumption, people freely produce content for corporations to profit from and generate atomizing imagery that perpetuates extractive thinking at scale. TikTok and Instagram's algorithms prioritize images representing flawless white skin, perfectly toned bodies, and suburban-bourgeoisie lifestyles, while drowning users in advertisements that help them realize these white supremacist ideals. Ads for makeups, anti-aging regimens, athleisure, flood people's feeds because they encourage the ideals of white supremacy that maintain optimization, homogenization, and consumption. This is most evident in the cycles of consumption encouraged by the fast fashion industry. Influencers and celebrities post photos of their outfits, and consumers buy second-rate versions made to emulate their style at the moment. This feedback loop creates constant demand for new styles and encourages consumers to buy clothes more frequently to maintain their optimizing image, all while producing literal boatloads of garbage that ultimately occupies the land and pollutes the air in countries where more brown people live.

To scale the production and dissemination of atomizing imagery, artificial technology increasingly re-aestheticizes the world it represents so that the content that people produce and consume online increasingly serves as the medium in which they identify themselves. Over time, the constant imbalancing brought on by atomizing imagery leads to a complete disintegration of oneself from their body and an eroded appreciation for human individuality, through which human life is intrinsically valuable. Instead, people come to value their life as contingent on their alignment with atomizing imagery, which through its widespread consumption flattens and homogenizes  human identity. People reappraise their lives in order to optimize online engagement with their images and increase their second-order vitality. The irony is that optimizing one's image for maximal consumption dulls the contours of one's identity, and breeds mid identities, or simply being mid. Social media is flush with cursed images of people's lives, relationships, families, outfits, food, vacations, work, all of which operate as propaganda for the self to convince others they've obtained the very best and have arrived. Sadly, people believe they must embody those images or have others desire their images to feel complete or at least legible. However, obtaining an image doesn't lead to any sense of peace or self-actualization, and rather dissolves identity altogether. Instead of securing the individuality we desire, seeking and producing images is bound to depress us because it can only deliver the mediocrity we revile. Chasing images compels atomized bodies to forget that their existence matters beyond maintaining an aggrandized self and hyper-curated brand.
Self-branding pervades nearly all walks of life. As well as on social media, success in art markets and scientific disciplines are all about curating a self-image that communicates how refined, esoteric, or deepcut one's taste is against all others. In an age with excessive technical expertise, taste is a proxy for authenticity and trumps having skills or actually creating something of value. How often does an artist, scientist, or technocrat ask themselves why they're doing what they're doing? Do we need fruit juicers on the internet-of-things? Do we need thousands of research articles on social psychology? Do we need another artist positioning three paintings in a gallery while unhoused people die on the streets outside those walls? 

How often are we honest with ourselves about the purpose of our work and our contributions to society? Rarely ever, because what motivates most of this labor is the pursuit of an emboldened self and the optimization of one's social networks. These class traitors strive to maintain their influence at the top of their highly-constructed kingdom rather than address the practical needs of their community. Artists, writers, scientists, and the like now seek microcelebrity status in their highly specialized echochambers - the prominent thinkers and creators of our generation are more akin to influencers than stewards of a more equitable and just society. The model of social media runs deep and is degrading all corners of life. Stepping outside of this disgraceful paradigm will require that people question the drivers of their work and thinking. It will require creation and production without the desire of fame and recognition. Transformation will require labor that addresses practical questions, rather than maintaining self-serving and highly-refined personal endeavors.

We're all hungry for other people's attention; it's a basic human instinct. But if we continue seeking others' attention at scale, we'll experience increasing levels of reflective anxiety and dread. We will need to secure more and more attention to replace the void left from forgetting that we have our own, infinite source of attention. We won't escape death and become immortal seeking second-order vitality and optimizing our identity online, we will only help uphold \textit{empirical} expansion. In producing and consuming images in the hope to balance us, all we will do is provide free content that keeps social media companies floating along as they extract our most vital resources and tear us apart at our cores. We need a radical transformation to restructure our lives around our collective vitality. 

\section*{Break free from atomizing imagery}

How can we step out of the frame set by atomizing imagery? We search for a method or cure, because our insatiable pursuit of images presupposes that the emancipation and actualization we seek lies outside of us. Any solution based on images is atomizing and therefore alienating; the only path towards transformation is to rediscover our ability to attend to the world and others and to ultimately embody our collective, rather than individualistic, agency. 
When you completely attend to something – a flower, a person, a pain – is there a time interval between you and the object of your attention? What happens when you categorize an object as a type of flower, person, or pain? Your memory of those categories conditions your experience of those things right now. Memory pulls bodies into time and away from attention to categorize the world. We see a flower and grasp its vibrance only in relation to the vibrance of other flowers stored in our memory. We consider each person in light of our earlier decisions about others. We experience pain's sting in the context of our pleasant sensations yesterday. All of these representations are of the past and outside of attention, and comparing our experience in relation to what is stored in our memory creates a time-interval, which separates the attending body from its experience. 
Have you ever had a conversation with a friend and were contemplating your response while they were speaking? Were you able to attend to what they were saying while doing this? No. You were optimizing your response, and placed yourself outside of the moment (shared attention with your friend) by engaging in self-directed thought. All language (narratives) and all images (categories and cause-effect relationships) move the body out of the present moment in this way, by placing attention either towards the past or to the future. This orienting power is the separating force of narratives and images, and is why any image of Vitality (or narrative/pictorial based explanation of its mechanisms for that matter) positions actualization outside of actuality. Because images place actualization beyond the present, actualization based on images can never be obtained because there is only the present, and images are only of time.
We will never experience our unity giving atomizing imagery our attention. Fear is insurmountable in a reflective space, because you cannot escape the death of your memory. Attempting to do so depletes us all of our vitality. We are drowning in a deluge of othering and isolating narratives that position people against one another, because the dissonance is turned into profit. 
Technology companies promised an easier, better connected, more pleasurable life through the adoption of their solutions. But what has it actually delivered? Capitalist innovation has only made us more isolated, angry, and sick of it all. Most companies in the information sector fail to solve practical problems facing our lives, and rather promote the mandate of increased efficiency and optimization. As many of the technology companies of the early naughts face downsizing or collapse, the mandated growth of capitalism is forcing the technology sector to look for new avenues of profit. In the coming decades, they will move towards extracting new domains of resources by more precisely commodifying our attention and lives. This efficiency will be made possible by new forms of artificial technology. To sustain themselves, they will require increasing addiction to their systems and images, and will eagerly develop new forms of artificial technology to monetize our lives with increased precision and efficiency. Our lives will be increasingly optimized, and they will continue to exploit atomizing imagery and narratives to do so. 

Those at the helms of artificial technology set the collective goals of our society by feeding us the stories they choose, narratives that instill hierarchy and violence towards one another. They spin news narratives that divide and distract us, convincing people that the goal is to have one political side dominate the other. Tethered to our constructed camps, believing in separating narratives, we lose track of the main point: those at the top of the systems generating atomizing imagery are the source of strife and violence, not others lacking agency in this collapsing system. Only once we overcome the binaries imposed by our imperialist overlords can we take back our attention, understand the other, and truly transform.

Transformation requires stepping away from senseless development. Development only creates more chaos and confusion. Images are based on images of other images which only steer us farther from the truth. Development consumes our lives with separating images and we've forgotten the simple fact that one's vitality is another's vitality. There is only our shared attention, and it is through our shared attention that we can express our collective agency and can completely transform and overcome the confusion brought on by atomizing imagery, media of violence, state, hierarchy, and empires.

How can we dismantle the vast networks of artificial technology that sustain the production of atomizing imagery? To be destructive, one must step outside of the fold to see the air. One must understand that engagement metrics are no substitute for actual experience; and the only way to gain agency over one's attention is to relegate images to their rightful status as mere and incomplete proxies for the sliver of the world that they represent. Atomizing imagery prompts acts of possession and ownership, and we must see within ourselves how we are possessive and seek to own others. Our need for security obscures the lens that produces a binary that breeds an infinite space within. Beyond the grasp of senseless categories and causal relations, one experiences Vitality and the space between all things collapses. Vitality is very quiet, and the cacophony of stories we tell ourselves drowns us so we can never fully experience our proclivity for change. 

\bibliographystyle{apalike}
\bibliography{bib}

\begin{thebibliography}{}

\bibitem[Banks, 2022]{banks2022attention}
Banks, D.~A. (2022).
\newblock The attention economy of authentic cities: How cities behave like influencers.
\newblock {\em European planning studies}, 30(1):195--209.

\bibitem[Csikszentmihalyi, 1975]{noauthor_play_1975}
Csikszentmihalyi, M. (1975).
\newblock Play and {Intrinsic} {Rewards}.
\newblock {\em Journal of Humanistic Psychology}, 15(3):41--63.
\newblock Publisher: SAGE Publications Inc.

\bibitem[Hacking, 1996]{hacking1996looping}
Hacking, I. (1996).
\newblock {The looping effects of human kinds}.
\newblock In {\em {Causal Cognition: A Multidisciplinary Debate}}. Oxford University Press.

\bibitem[Hurley, 2023]{hurley2023conspiracist}
Hurley, R. (2023).
\newblock {\em Conspiracist Manifesto}.
\newblock Semiotext(e), MIT Press, Los Angeles.

\bibitem[Tiqqun, 2010]{tiqqun2010introduction}
Tiqqun (2010).
\newblock {\em Introduction to Civil War}.
\newblock Semiotext(e), MIT Press, Los Angeles.

\bibitem[Tiqqun, 2012]{tiqqun2012toyg}
Tiqqun (2012).
\newblock {\em Preliminary Materials for Theory of the Young-Girl}.
\newblock Semiotext(e), MIT Press, Los Angeles.

\bibitem[Weil, 1997]{weil1997gravity}
Weil, S. (1997).
\newblock {\em Gravity and Grace}.
\newblock U of Nebraska Press.

\end{thebibliography}

\end{multicols}

\end{document}